\documentclass[prl,aps,twocolumn,superscriptaddress]{revtex4-1}
\usepackage{amssymb,amsfonts,amsmath,amstext,graphicx,hyperref,lipsum}
\usepackage[normalem]{ulem}
\hypersetup{
	colorlinks=true,
	linkcolor=blue,
	filecolor=magenta,      
	urlcolor=cyan,
}
\usepackage[dvipsnames]{xcolor}
\usepackage{tikz}
\usepackage{pgfkeys}
\usetikzlibrary{shapes.callouts}
\tikzset{
  level/.style   = { ultra thick, blue },
  connect/.style = { dashed, red },
  notice/.style  = { draw, rectangle callout, callout relative pointer={#1} },
  label/.style   = { text width=2cm }
}
\newcommand{\tr}{\mbox{tr}}
\newcommand{\rem}[1]{}
\newcommand{\add}[1]{{\color{black}#1}}

\def\tr{\mbox{tr}}
\def\bra#1{\langle{#1}|}
\def\ket#1{|{#1}\rangle}

{\catcode`\|=\active
  \gdef\Braket#1{\begingroup
\mathcode`\|32768\let|\BraVert\left<{#1}\right>\endgroup}}
\def\BraVert{\egroup\,\mid\,\bgroup}

\begin{document}

\title{Quantum Synchronization in Nanoscale Heat Engines}

\author{Noufal Jaseem}
\email{noufal@iitb.ac.in}
\affiliation{Department of Physics, Indian Institute of Technology-Bombay, Powai, Mumbai 400076, India}

\author{Michal Hajdu\v{s}ek}
\email{cqtmich@nus.edu.sg}
\affiliation{Centre for Quantum Technologies, National University of Singapore, 3 Science Drive 2, 117543 Singapore, Singapore}

\author{Vlatko Vedral}
\affiliation{Centre for Quantum Technologies, National University of Singapore, 3 Science Drive 2, 117543 Singapore, Singapore}
\affiliation{Department of Physics, University of Oxford, Parks Road, Oxford, OX1 3PU, UK}

\author{Rosario Fazio}
\affiliation{Centre for Quantum Technologies, National University of Singapore, 3 Science Drive 2, 117543 Singapore, Singapore}
\affiliation{ICTP, Strada Costiera 11, 34151 Trieste, Italy}
\affiliation{NEST, Scuola Normale Superiore \& Instituto Nanoscienze-CNR, I-56126 Pisa, Italy}

\author{Leong-Chuan Kwek}
\affiliation{Centre for Quantum Technologies, National University of Singapore, 3 Science Drive 2, 117543 Singapore, Singapore}
\affiliation{Institute of Advanced Studies, Nanyang Technological University, Singapore 639673}
\affiliation{National Institute of Education, Nanyang Technological University, Singapore 637616}

\author{Sai Vinjanampathy}
\email{sai@phy.iitb.ac.in}
\affiliation{Department of Physics, Indian Institute of Technology-Bombay, Powai, Mumbai 400076, India}
\affiliation{Centre for Quantum Technologies, National University of Singapore, 3 Science Drive 2, 117543 Singapore, Singapore}

\date{\today}

\begin{abstract}
	Owing to the ubiquity of synchronization in the classical world, it is interesting to study {its behavior} in quantum \rem{regime} \textcolor{black}{systems}. Though quantum synchronisation has been investigated in many systems, a clear connection to quantum technology applications is lacking.
	\rem{Hence a clear experimental demonstration of quantum synchronization is lacking.} We  bridge this gap and show \rem{how} \textcolor{black}{that nanoscale heat engines are a natural platform to study quantum synchronization and always possess a stable limit cycle.}
	\rem{We demonstrate how the power output of the quantum engine shows signatures of synchronization which can be readily observed with current technologies.}
	\textcolor{black}{Furthermore, we demonstrate an intimate relationship between the power of a heat engine and its phase-locking properties by proving that synchronization places an upper bound on the achievable steady-state power of the engine.}
	\textcolor{black}{Finally, we show that the efficiency of the engine sets a point in terms of the bath temperatures where synchronization vanishes.}
	We link the physical phenomenon of synchronization with the emerging field of quantum thermodynamics by establishing quantum synchronization as a mechanism of stable phase coherence. 
\end{abstract} 
\maketitle

\makeatletter

\emph{Introduction.---}
\textcolor{black}{Synchronization has been observed and studied in a multitude of naturally occurring as well as man-made systems, finding different applications in fields ranging from engineering to medicine \cite{pikovsky2003synchronization, kuramoto2012chemical, balanov2008synchronization, strogatz2004sync}.} Recently the universality of this phenomenon has also been embraced in the quantum realm. Thus far all the efforts have been aimed at characterising synchronization in quantum systems\cite{heinrich2011collective,bastidas2015quantum,witthaut2017classical,marquardt2006dynamical, lee2013quantum, walter2014quantum,amitai2017synchronization,sonar2018squeezing,sonar2018squeezing,lorch2017quantum,nigg2018observing,lorch2017quantum,roulet2018synchronizing,roulet2018quantum,kwek_vp}.
It is compelling to ask if/where this phenomenon may play a role in the functionality itself of quantum devices.
What is the impact of synchronization on quantum technology platforms?
A natural playground to  explore this question is quantum thermodynamics~\cite{goold2016role,vinjanampathy2016quantum}.
Nanoscale heat engines can be modelled as  multi-level atoms that  produce work when cyclically coupled to two or more heat baths~\cite{kosloff2013quantum}.
In this manuscript we highlight a deep connection between quantum synchronization and the performance \rem{in} \textcolor{black}{of} nanoscale heat engines.
This relation is particularly evident in the emitted power of the engine displaying an Arnold tongue, a distinct signature of synchronization.
\textcolor{black}{We show that synchronization provides an upper bound on the achievable magnitude of the steady-state power of the heat engine.
Finally, we determine that the efficiency of the heat engine $\eta$ sets the ratio of bath temperatures where synchronization vanishes.
This points at a fundamental relation between the operational regime of a thermal machine and its synchronization properties.}
 An important outcome of this relationship is to bring the study of quantum synchronisation closer to quantum technology applications.

\begin{figure*}[t]
	\includegraphics[width=0.8\textwidth]{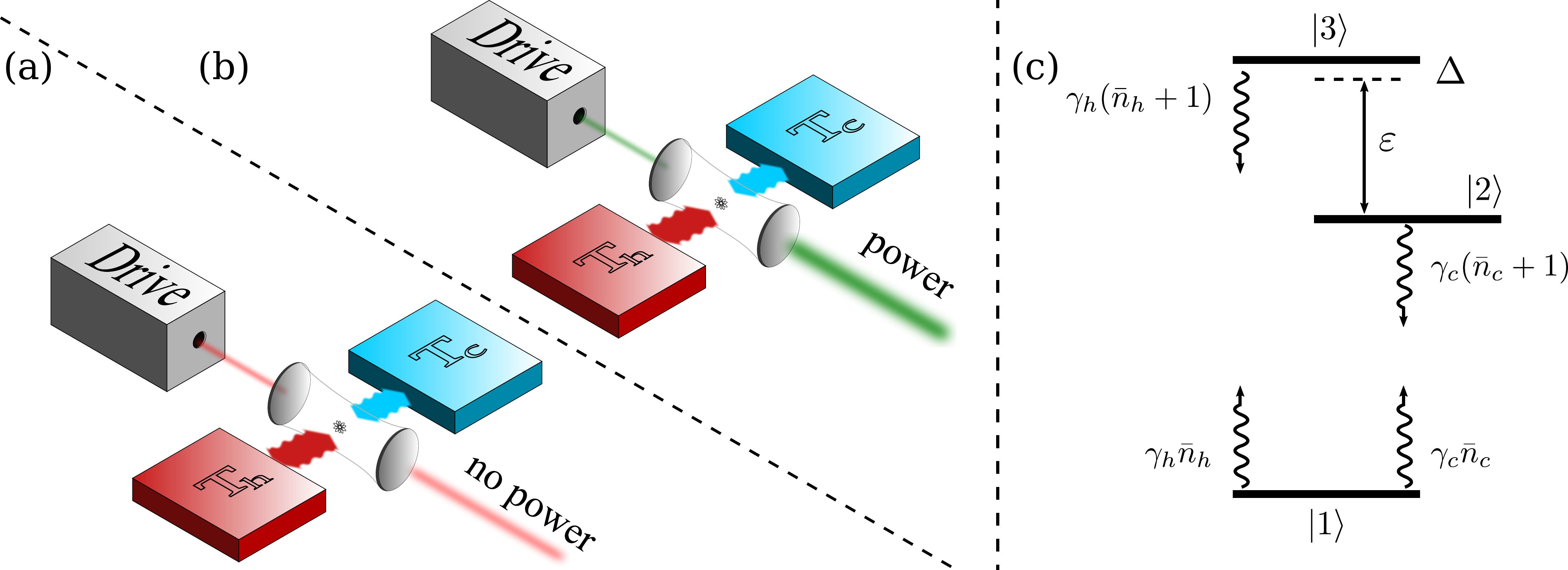}
	\caption{\label{sync_engine} A quantum thermal machine can be used to observe quantum synchronization. The engine consists of a three level maser driven three-level atom which generates output power when coupled to two dissimilar baths. The relationship between power, detuning and the driving strength is understood as arising from the synchronization of the three-level atom to the external drive. At a given driving strength, if the driving field is far detuned from the relevant maser transition, the system cannot synchronize to the external drive. The output power is very low as seen in (a). If the detuning is in the synchronization region, the engine power is reinforced by synchronization as depicted in (b). (c) Transition between levels $\ket{2}$ and $\ket{3}$ is coupled by a coherent field of strength $\varepsilon$. Levels $\ket{1}$-$\ket{3}$ are coupled by a hot bath at temperature $T_h$ while levels $\ket{1}$-$\ket{2}$ are coupled by a cold bath at temperature $T_c$.}
	\label{engine}
\end{figure*}

A basic prerequisite for synchronization is the existence of a stable limit cycle \cite{pikovsky2003synchronization}.
Once a limit cycle has been established in a non-linear dynamical system, we can synchronize the given system to \rem{either} an external frequency standard \footnote{\textit{synchronisation} is usually reserved for the adjustment of rhythms between two or more physical systems, whereas the adjustment of a systems rhythm to an external frequency is known as \textit{entrainment}. Following existing literature in quantum synchronisation, we will use the same word for both phenomena.}.
We show that thermal atoms possess a limit cycle by constructing the quasi-probability distribution for arbitrary three-level atoms.
This allows us to perform a synchronization analysis of a thermal three-level system and investigate its phase-locking properties in the context of the engine's performance. 
\rem{We then investigate phase locking of thermal atoms to an external field and show how the power spectral density of the emitted radiation exhibits signatures of synchronization.}
\rem{Finally, we show an intimate relationship between synchronization and the performance of thermal three-level masers as a heat engine.}

\emph{Thermal three-level atoms.---}
We consider the  three-level maser model introduced by Scovil and Schulz-Dubois as an example of quantum synchronization \cite{scovil1959three}.
In view of the connection to  heat engines, we will examine its properties when coupled to two baths kept at different temperatures.
Though we consider a specific model to develop the results presented here, the techniques are transferrable to generic coherent quantum technologies.

\rem{We consider thermal three-level atoms to study synchronization.}
The evolution for the system depicted in Fig.~(\ref{engine}) can be written ($\hbar=1$) as
\begin{align}\label{lindy}
    \dot{\rho}=-i[H,\rho]+\mathcal{L}_h[\rho]+\mathcal{L}_c[\rho],
\end{align}
where $H=H_0+V$ represents the sum of the bare Hamiltonian $H_0$ and the drive $V=\varepsilon(e^{i\omega_d t}\sigma_{23}+e^{-i\omega_d t}\sigma_{32})$, with $\sigma_{ij}\equiv\ket{i}\bra{j}$. We model the baths as single-mode thermal fields resonant with the respective atomic transitions that they couple.
The bath frequency $\omega$, its temperature $T$ and its average photon number $\bar{n}$ are related by $e^{-\hbar\omega/k_B T} = \bar{n}/(\bar{n} + 1)$ \cite{breuer2002theory}.
The Lindbladian dissipators are given by $\mathcal{L}_h[\rho]\equiv \gamma_h \bar{n}_h \mathcal{D}[\sigma_{31}]\rho + \gamma_h (1+\bar{n}_h) \mathcal{D}[\sigma_{13}] \rho$, \textcolor{black}{incoherently coupling the $\ket{1}\leftrightarrow\ket{3}$ transition at temperature $T_h$}, and $\mathcal{L}_c[\rho]\equiv\gamma_c \bar{n}_c \mathcal{D}[\sigma_{21}] \rho+ \gamma_c  (1+\bar{n}_c) \mathcal{D}[\sigma_{12}]\rho$ \textcolor{black}{representing a cold bath at temperature $T_c$ and incoherently coupling the $\ket{1}\leftrightarrow\ket{2}$ transition}.
The Markovian master equation is written in the standard Lindblad form \cite{breuer2002theory} as $\mathcal{D}[O]\rho=O\rho O^{\dag}-\frac{1}{2}\{O^{\dag}O,\rho\}$.

The system is weakly driven with small detunings, which ensures that (a) the limit cycles presented below are not deformed by the drive, (b) the populations in the steady states are dominated by the dissipation rates and (c) the adiabatic definitions of heat and work are valid in this regime.
\textcolor{black}{
Points (a) and (b) are related and ensure that we are in a regime where it is sensible to study synchronization \cite{roulet2018synchronizing}.
Large driving strength also produces phase locking.
However, due to deformation of the limit cycle this regime is known as suppression of natural dynamics \cite{balanov2008synchronization} and is beyond the scope of this manuscript.
In general, the power of an engine can be attributed to two parts, one that accounts for the change in the instantaneous eigenvalues and another which accounts for the change in eigenvectors.
For perturbative driving the first term does not contribute (indeed our eigenvalues are constant in time).}

\rem{The definition of heat ($\delta Q=\tr\{ H\delta\rho\}$) and work ($\delta W=\tr\{\delta H\rho\}$) due to Alicki \cite{alicki1979quantum} are known to (a) not saturate in steady state and (b) be frame dependent quantities.
This motivated Boukobza and Tannor to define heat \textcolor{black}{current as $\dot{Q} =\text{tr}\left\{ \mathcal{L}[\rho] H_0\right\}$ and power as $P = -i \; \text{tr}\left\{ [H, \rho] H_0 \right\}$}~\cite{boukobza2006thermodynamics,boukobza2007three}, which are known to behave appropriately.
The analytical steady state of this system, $\rho^{ss}$ can be calculated and is discussed in the supplemental material alongside formulae for \textcolor{black}{heat currents and power} for completeness.
We note that since our driving is perturbative, the two definitions only differ by a small number of the order of $\varepsilon$.}

\begin{figure*}[t]
\includegraphics[width=0.98\textwidth]{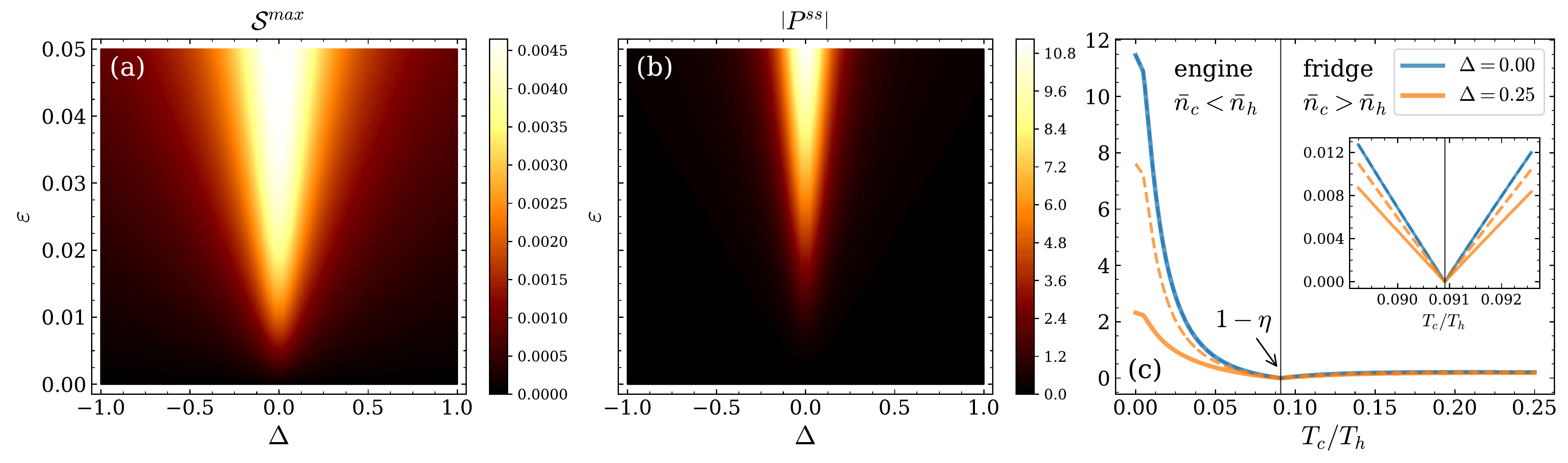}
\caption{\label{fig:contours} \textcolor{black}{(a) Synchronization measure $\mathcal{S}^{max}$ for different detuning $\Delta$ and driving strength $\varepsilon$ displaying the Arnold tongue. (b) Magnitude of the steady-state power $|P^{ss}|$ as a function of $\Delta$ and $\varepsilon$. The remaining parameters for both plots are fixed at $\gamma_h=10^{-2}$, $\bar{n}_h = 5$, $\bar{n}_c= 10^{-3}$, $\gamma_c=10 \gamma_h$, and $\omega_{32}=10/\gamma_h$. (c) Comparison of power $|P^{ss}|$ (solid lines) with the synchronization bound $16\pi\vert\varepsilon\vert \omega_{32}\mathcal{S}^{max}$ (dashed lines) as functions of the ratio of bath temperatures $T_c/T_h$. For resonant driving, $\Delta=0$, the two curves coincide while for $\Delta=0.25$ we see the synchronization upper bounds $|P^{ss}|$. The black vertical line marks $T_c/T_h=\omega_{21}/\omega_{31}$ where $\bar{n}_c=\bar{n}_h$ and is related to the efficiency of the engine $\eta$ via $\omega_{21}/\omega_{31}=1-\eta$.  At this point the synchronization bound vanishes and therefore does $|P^{ss}|$. For $\bar{n}_c<\bar{n}_h$ the three-level system acts as an engine while for $\bar{n}_c>\bar{n}_h$ it acts as a fridge. The driving strength is fixed at $\epsilon=0.05$ and $\omega_{21}=1/\gamma_h$. The inset shows that the synchronization bound and therefore $|P^{ss}|$ both vanish at the point where $\bar{n}_c=\bar{n}_h$.}}
\end{figure*}

\emph{Synchronization in heat engines.---} A stable limit cycle is defined as a phase trajectory that attracts nearby orbits.
Such a limit cycle arises as a compromise between dissipation and gain in the system.
Another defining property of the limit cycle is that an observable phase is free.
It is this freedom of the phase combined with the stability of the limit cycle that allows a weak external signal to influence the phase distribution of the oscillator.
To study limit cycles in thermal three-level systems, we define the Husimi-Kano $Q$-representation function \cite{schleich2011quantum}, given for a generic three-level atom as \textcolor{black}{$Q(\theta,\xi,\varphi_1,\varphi_2)\equiv(6/\pi^2)\bra{n_3}\rho \ket{n_3}$} where $\ket{n_3}\equiv(\cos \theta,\; e^{i \varphi_1}\cos \xi \sin \theta,\; e^{i \varphi_2}\sin \xi \sin \theta)^T$ is the SU(3) coherent state \cite{nemoto2000generalized}.
\textcolor{black}{
From the definition of $\ket{n_3}$ we can see that $(\varphi_1,\varphi_2)$ and $(\theta,\xi)$ carry information about the coherences and the populations of the system, respectively.}
Under the influence of a bare Hamiltonian written in the energy eigenbasis, the SU(3) coherent state evolves as \textcolor{black}{$\ket{n_3}=\ket{\theta,\xi,\varphi_1,\varphi_2}\rightarrow\ket{\theta,\xi,\varphi_1-\omega_{21}t,\varphi_2-\omega_{31}t}$, where $\omega_{ij}\equiv\omega_i-\omega_j$}.
This implies that the angles $(\varphi_1,\varphi_2)$ are the relevant dynamical phases to the study of synchronization.
\textcolor{black}{
Similarly to a spin-1 atom considered in \cite{roulet2018synchronizing} the system becomes synchronized when the distribution of $\varphi_1$ and $\varphi_2$ becomes localized.
The difference between the localized and uniform distributions serves as a measure of synchronization,} 
\begin{equation}
 	\mathcal{S}(\varphi_1,\varphi_2) \equiv \int d\Omega~Q(\theta,\xi,\varphi_1,\varphi_2) - \frac{1}{4\pi^2},
  	\label{eq:S_measure_def}
\end{equation}
where $\int d\Omega\equiv\int_0^\frac{\pi}{2} d\theta \cos \theta \sin^3 \theta \int_0^\frac{\pi}{2} d\xi \cos \xi \sin \xi$.
\rem{The measure $\mathcal{S}(\varphi_1,\varphi_2)$ quantifies the deviation of the phase distribution from the uniform distribution and hence measures the phase locking of the angles $(\varphi_1,\varphi_2)$.}
This is a generalisation to the full SU(3) group of a previous definition presented for SO(3) subgroup of SU(3) \cite{roulet2018synchronizing}.

Under the influence of the bare Hamiltonian and the dissipators, the system settles into a stable limit cycle given by no phase preference in the angles $(\varphi_1,\varphi_2)$.
When the system is driven, it develops a phase preference characterised by localization of $Q(\theta,\xi,\varphi_1,\varphi_2)$ in the relevant angles $(\varphi_1,\varphi_2)$.
\textcolor{black}{The synchronization measure $\mathcal{S}(\varphi_1,\varphi_2)$ in the steady state is evaluated to be}
\begin{eqnarray}
	\mathcal{S}(\varphi_1,\varphi_2) &=& \frac{1}{8\pi} \bigg \{ \text{Re} \left[ e^{i\varphi_1}\rho^{ss}_{12} \right] + \text{Re} \left[ e^{i\varphi_2}\rho^{ss}_{13} \right] \nonumber \\
	&+& \text{Re} \left[ e^{i(\varphi_2-\varphi_1)}\rho^{ss}_{23} \right] \bigg \}.
	\label{eq:S_measure}
\end{eqnarray}
Here $\rho^{ss}$ is the steady-state density matrix.
For the evolution given in Eq.~(\ref{lindy}), this synchronization measure \textcolor{black}{simplifies to $\mathcal{S}(\varphi_1,\varphi_2)=\text{Re} \left[ e^{i(\varphi_2-\varphi_1)}\rho_{23} \right]/8\pi$.
We are primarily concerned with how much the phases are localised in the steady state, given by the maximum of $\mathcal{S}(\varphi_1,\varphi_2)$ which can be readily computed, 
\begin{equation}
	\mathcal{S}^{max} \equiv \max_{\varphi_1,\varphi_2} \mathcal{S}(\varphi_1, \varphi_2) = \frac{1}{16\pi}C_{l_1}(\rho^{ss}),
	\label{eq:S_max}
\end{equation}
where $C_{l_1}(\rho)\equiv\sum_{i\neq j}|\rho_{ij}|$ is the $l_1$-norm of coherence \cite{baumgratz2014quantifying}.
\rem{We note that Eq.~(\ref{eq:S_max}) holds for the evolution given by Eq.~(\ref{lindy}) and for more general evolution the right-hand-side of Eq.~(\ref{eq:S_max}) becomes an upper bound on $\mathcal{S}^{max}$.}
The synchronization measure $\mathcal{S}^{max}$ is shown in Fig.~\ref{fig:contours}(a) as a function of the drive's detuning $\Delta=\omega_{32}-\omega_d$ and its strength $\varepsilon$.}
In analogy to classical synchronization, note the Arnold tongue-like behaviour of the figure indicating the phase locking of the atom to the external drive.
The range of detuning for which the atom displays phase locking to the external drive increases with the strength of the drive.

\emph{Engine Performance \& Power Arnold tongue.---} Efficiency and power are the characteristics that most often define the operation of an engine \cite{millen2016perspective,goold2016role,vinjanampathy2016quantum}.
Scovil and Schulz-Dubois \cite{scovil1959three} showed that a maser can be operated as a heat engine whose efficiency is bounded by Carnot efficiency, namely $\eta={\omega_{32}}/{\omega_{31}}\leq\eta_c\equiv1-{T_c}/{T_h}$.
\textcolor{black}{This bound justifies our choice of driving a generic three-level atom, whose Husimi-Kano $Q$ function is defined in terms of SU(3) coherent states as opposed to considering an equally spaced three-level atom of \cite{roulet2018synchronizing}.
Our generalisation reveals the complex dynamics inherent to the full SU(3) group while simultaneously allowing the efficiency to be varied by the choice of $\omega_{ij}$.}

A natural question arises, namely if the aforementioned thermodynamical characteristics depend on the underlying dynamical properties of the system, undergoing a transition from synchronized to unsynchronized regime.
The efficiency only depends on the transition frequencies of the system and is independent of any dynamical properties.
This is because the efficiency tracks the energy transactions from the bath to the piston via the working medium, and is unconcerned about the dynamical processes involved \textcolor{black}{in the perturbative driving regime}. 

On the other hand, the power of an engine is intimately related with dynamics \cite{kosloff2013quantum}.
To operate an engine at the Carnot efficiency, one must perform adiabatic strokes which will not generate any heat.
Since adiabatic strokes take infinite time, such an engine does not generate power.
Likewise in the quantum regime, adiabatic strokes are transitionless and do not generate any excitations, allowing the engine to follow the instantaneous eigenstate of the total Hamiltonian.
Adiabatic driving has to thermalize with the bath at each instant in time, also implying the slowing down of dynamics \cite{suri2017speeding,dann2018shortcut}. \add{Furthermore, perturbative driving does not change eigenvalues, and hence the contribution to power cannot arise from such terms.}
Such a scenario hence precludes generating power from diagonal density matrices. To generate power, the density matrix hence has to be off-diagonal in the Hamiltonian eigenbasis \cite{feldmann2003quantum}. \add{Any definition of power has to satisfy all these constraints and has to be frame independent, saturate in the steady state and satisfy the second law of thermodynamics for all values of detuning $\Delta$ unlike the definition presented in \cite{alicki1979quantum}.}
This intuition is captured by the steady-state power $P^{ss}=-i \; \text{tr}\{[H, \rho^{ss}]H_0\}$ due to Tannor and Boukobza \cite{boukobza2006thermodynamics,boukobza2007three} \add{and satisfies all the conditions stated above.}
A detailed analysis of the definitions of heat and work, alongside the engine performance is presented in the \textcolor{black}{supplemental material} for completeness.
\rem{Our drive is perturbative, since entrainment (synchronization) is best studied as an adjustment of rhythms that happens in the presence of perturbative driving (coupling).
Likewise the thermodynamics of driven systems is uncontroversial in the perturbative regime.
In general, the power of an engine can be attributed to two parts, one that accounts for the change in the instantaneous eigenvalues and another which accounts for the change in eigenvectors.
Since we are driving our quantum system perturbatively, the first term does not contribute (indeed our eigenvalues are constant in time).}

The magnitude of steady-state power $|P^{ss}|$ is presented in Fig.~\ref{fig:contours}(b) as a function of $\Delta$ and $\varepsilon$, and displays an Arnold tongue-like behaviour.
To understand this, we note that the power is non-zero only when the steady state density matrix is off-diagonal with respect to the bare Hamiltonian.
This means that to observe a powerful quantum engine, the system has to be driven externally by a Hamiltonian that does not commute with the bare Hamiltonian.
Power is only produced in the region where we observe phase locking to the external drive.
If the system is sufficiently detuned at a given value of the driving strength $\varepsilon$, negligible power is emitted by the maser.
The underlying mechanism for this Arnold tongue-like behaviour is synchronization which produces coherence in the Hamiltonian eigenbasis.

\textcolor{black}{To make this connection between synchronization and power quantitative we begin by calculating the steady-state power explicitly,
\begin{equation}
	P^{ss} = 2 \varepsilon \omega_{32} \text{Im} \left[ \rho^{ss}_{23} \right].
	\label{eq:power}
\end{equation}
Using $\left| \text{Im}\left[ \rho^{ss}_{23} \right] \right| \leq \left| \rho^{ss}_{23} \right|$ along with Eq.~(\ref{eq:S_max}) leads to a synchronization bound on the magnitude of the steady-state power of the engine,
\begin{equation}
	\left| P^{ss} \right| \leq 16 \pi \left| \varepsilon \right| \omega_{32} \mathcal{S}^{max}.
	\label{eq:power_vs_Smax}
\end{equation}
The bound becomes an equality when the engine is driven resonantly, $\Delta=0$, because in this case $\rho^{ss}_{23}$ becomes pure imaginary.
This can be seen by computing $\rho_{23}^{ss}$,
\begin{equation}
	\rho_{23}^{ss} = i \varepsilon \Gamma_{23}^* \gamma_c \gamma_h \left( \bar{n}_c - \bar{n}_h \right) / \beta,
	\label{eq:rho23}
\end{equation}
where $\Gamma_{23} = [\gamma_h (\bar{n}_h + 1) + \gamma_c (\bar{n}_c + 1)]/2 - i\Delta$ and $\beta$ is a real number whose exact form is given in the supplemental material.
Inequality (\ref{eq:power_vs_Smax}) is our main result.
It directly links the power of a heat engine with the amount of synchronization in the system.
Synchronization sets an upper bound on the achievable power of an engine for a fixed driving strength $\varepsilon$.}

We explore this connection further and study how the synchronization bound in (\ref{eq:power_vs_Smax}) varies with bath temperatures. Fig.~\ref{fig:contours}(c) shows the magnitude of the steady-state power $|P^{ss}|$ (solid lines) as a function of the ratio of bath temperatures $T_c/T_h$, controlled by varying $\bar{n}_c$ and keeping $\bar{n}_h$ fixed.
Strong power is achieved for vanishing $T_c/T_h$ which is expected and then decreases as $T_c/T_h$ increases.
We see that $|P^{ss}|$ saturates the synchronization bound (dashed lines) when the engine is driven resonantly while in the case of finite detuning $|P^{ss}|$ is lower than the synchronization bound.
As $T_c/T_h$ increases both $|P^{ss}|$ and the synchronization bound decrease until we reach an interesting point where $\bar{n}_c=\bar{n}_h$ and the system stops behaving like an engine.
It can be shown that this point corresponds to $T_c/T_h=\omega_{21}/\omega_{31}=1-\eta$.
This shows that the engine efficiency $\eta$ determines the point where the synchronization measure $\mathcal{S}^{max}$ and therefore $|P^{ss}|$ vanish as can be seen from Eq.~(\ref{eq:rho23}) and is also shown in the inset of Fig.~\ref{fig:contours}(c).
As the temperature ratio increases past $T_c/T_h=1-\eta$ the synchronization bound becomes finite again and so does $|P^{ss}|$, and the system enters a new regime where it behaves like a fridge, where heat currents and power reverse sign and the energy starts flowing from the cold bath to the hot one as detailed in the supplemental material.
Note that the synchronization bound in (\ref{eq:power_vs_Smax}) holds true regardless of whether the system is operated as an engine or a fridge.

\emph{Discussion.---} 
We have demonstrated an intimate connection between synchronization and nanoscale heat engines and showed how the properties of the output power can be understood as a consequence of synchronization developing in the working fluid of the engine. 
An explicit example of this connection is the case of three-level atoms connected to two thermal reservoirs.
\textcolor{black}{We have derived an upper bound on the engine power in terms of the measure of synchronization, $\mathcal{S}^{max}$.
The maximum amount of synchronization as measured by $\mathcal{S}^{max}$ can be operationally understood as the maximum magnitude of steady-state power of a three-level engine when it is driven resonantly.
Finally, we have showed that the engine's efficiency $\eta$ determines the ratio of the bath temperatures where synchronization vanishes.} 

This connection offers, in addition, a very important route towards the experimental observation of quantum synchronization since masers are mature quantum platforms.
Three-level atoms have been experimentally coupled to thermal baths using magneto-optical traps \cite{zou2017quantum} pumped by incoherent laser beams.
We envisage however that the connection is far more general and it applies to any generic multi-level quantum engine.
We hence anticipate that signatures of quantum synchronization can be observed experimentally.
We note that the underlying principle of synchronization of multi-level systems with linear baths is universal and can be applied wherever a stable phase relationship between atomic levels is needed.
This will pave the way for future quantum technologies based on quantum synchronization.

\section*{acknowledgements}
Centre for Quantum Technologies is a Research Centre of Excellence funded by the Ministry of Education and the National Research Foundation of Singapore. This research is supported by the National Research Foundation, Prime Minister's Office, Singapore under its Competitive Research Programme (CRP Award No. NRF-CRP14-2014-02). SV also acknowledges support from an IITB-IRCC grant number 16IRCCSG019 and a DST-SERB Early Career Research Award (ECR/2018/000957).


%

\clearpage

\newpage 
\widetext

\section*{Supplementary Material}

\emph{Measure of synchronization.---}	
In this section we extend the synchronization measure to a general three-level system using SU(3) coherent states.
We follow the SU(3) coherent state analysis due to Nemoto~\cite{nemoto2000generalized} where the SU(3) coherent state is 
\begin{equation}
\arrowvert n_3 \rangle = (\cos \theta,\; e^{i \varphi_1}\cos \xi \sin \theta,\; e^{i \varphi_2}\sin \xi \sin \theta)^T,
\end{equation}
where $0\leq \theta, \xi\leq \pi/2$, and $0\leq \varphi_1, \varphi_2 \leq 2 \pi$.
The SU(3) group measure is given by~\cite{nemoto2000generalized}
\begin{equation}
dv = d\theta d\xi d\varphi_1 d\varphi_2\cos \theta \sin^3 \theta \cos \xi \sin \xi,
\end{equation}	
which leads to the completeness relation for the coherent state $\ket{n_3}$,
\begin{equation}
	 \int_0^{2\pi} d\varphi_1  \int_0^{2\pi} d\varphi_2\int_0^{\pi/2} d\theta \cos \theta \sin^3 \theta \int_0^{\pi/2} d\xi \cos \xi \sin \xi \ket{n_3}\bra{n_3}  = \frac{\pi^2}{6} \mathbb{I}.
\end{equation}
The steady state Husimi-Kano $Q$ function can be written as
\begin{eqnarray}
	Q^{ss} &=& \frac{6}{\pi^2} \bra{ n_3 } \rho^{ss} \ket{ n_3 } \nonumber\\
	&=&  \frac{6}{\pi^2} \left( \rho^{ss}_{33} \cos^2 \theta + \rho^{ss}_{22} \sin^2 \theta \cos^2 \xi + \rho^{ss}_{11} \sin^2 \theta \sin^2 \xi \right) \nonumber \\
	&+& \frac{6}{\pi^2} \left( \sin 2\theta \cos \xi \;Re[e^{i\varphi_1} \rho^{ss}_{12}]  + \sin 2\theta \sin \xi \;Re[e^{i\varphi_2} \rho^{ss}_{13}] + \sin^2 \theta  \sin 2 \xi \;Re[e^{i(\varphi_2-\varphi_1)} \rho^{ss}_{23}] \right)
 \label{eq:husimiSU3}
 \end{eqnarray}
 One can see that in the absence of off-diagonal elements, $Q^{ss}$ is independent of the phases $\varphi_1$ and $\varphi_2$, and hence the Eq.~(\ref{eq:husimiSU3}) establishes the limit cycle for a system evolving under thermal Lindbladians.
 
Under the action of the bare Hamiltonian,
\begin{equation}
H_0 =  \left(\begin{array}{ccc}
	\omega_1 & 0 & 0  \\
	0 &  \omega_2 & 0 \\
	0 & 0 & \omega_3
	\end{array}\right),
\end{equation}
 the coherent state evolves as
$e^{-i H_0 t} \arrowvert n_3(\theta,\xi, \varphi_1,\varphi_2)\rangle = \arrowvert n_3(\theta,\xi, \varphi_1 - \omega_{21} t,\varphi_2 - \omega_{31}t)\rangle$, where $\omega_{ij}\equiv\omega_i-\omega_j$.
There are two parameters, phases $\varphi_1$ and $\varphi_2$, that are evolving with time,  thus one has to look at the system's preference for both of these phases.
We can define the synchronization measure for a general three-level system as
\begin{eqnarray}
	\mathcal{S}(\varphi_1,\varphi_2) &=& \int_0^{\pi/2} d\theta \cos \theta \sin^3 \theta \int_0^{\pi/2} d\xi \cos \xi \sin \xi  \; Q^{ss}(\theta,\xi,\varphi_1,\varphi_2) - \frac{1}{4\pi^2} \nonumber\\
	&=& \frac{1}{8 \pi} \left\{ \text{Re}[e^{i\varphi_1} \rho^{ss}_{12} ]+ \text{Re}[e^{i\varphi_2} \rho^{ss}_{13} ] +  \text{Re}[ e^{i(\varphi_2-\varphi_1)} \rho^{ss}_{23} ]\right\}.
	\label{eq:GenSync}
\end{eqnarray}
We can now maximize $\mathcal{S}(\varphi_1,\varphi_2)$ over the two phases,
\begin{eqnarray}
	\mathcal{S}^{max} &=& \max_{\varphi_1,\varphi_2} \frac{1}{8\pi} \left\{ \text{Re} \left[ e^{i\varphi_1} \rho^{ss}_{12} \right] + \text{Re} \left[ e^{i\varphi_2} \rho^{ss}_{13} \right] + \text{Re} \left[ e^{i(\varphi_2 - \varphi_1)} \rho^{ss}_{23} \right] \right\} \nonumber \\
	&=& \max_{\varphi_1, \varphi_2} \frac{1}{8\pi} \left[ \cos(\varphi_1+\vartheta_{12}) |\rho^{ss}_{12}| + \cos(\varphi_2 + \vartheta_{13}) |\rho^{ss}_{13}| + \cos(\varphi_2 - \varphi_1 + \vartheta_{23}) |\rho^{ss}_{23}| \right] \nonumber \\
	&\leq& \frac{1}{8\pi} \left( |\rho^{ss}_{12}| + |\rho^{ss}_{13}| + |\rho^{ss}_{23}| \right) \nonumber \\
	&=& \frac{1}{16\pi} C_{l_1}(\rho^{ss}),
	\label{eq:GenSMax}
\end{eqnarray}
where we expressed the coherences in polar coordinates $\rho^{ss}_{ij}=e^{i\vartheta_{ij}}|\rho^{ss}_{ij}|$.
$\mathcal{S}^{max}$ saturates this bound only when $\vartheta_{23}=\vartheta_{13}-\vartheta_{12}$ or when two of the three coherences vanish.
In the main text, $\mathcal{S}(\varphi_1,\varphi_2)=\text{Re}[e^{i(\varphi_2 - \varphi_1)} \rho^{ss}_{23}] / 8\pi$ and $\mathcal{S}^{max}=C_{l_{1}}(\rho^{ss})/16\pi$ because two of the coherences vanish in the steady state, namely $\rho^{ss}_{12}=\rho^{ss}_{13}=0$.


\emph{Heat and Work for Quantum Engines.---}	
Let $\tilde{H}$ be the Hamiltonian in an arbitrary frame obtained by the unitary $U= e^{-i x t}$, ie. $\tilde{H} = {U}^\dagger H U$, where $H$ is the Hamiltonian in the Schr\"{o}dinger picture.
As the expectation values of an operator is independent of the frame, we have $\langle \tilde{H}\rangle=\langle H\rangle$.
By taking the derivative of $\langle \tilde{H} \rangle$ with respect to time, we get 
\begin{equation}
	\frac{d}{dt} \langle \tilde{H} \rangle = \text{tr}\left\{ \frac{d \tilde{H}}{dt} \tilde{\rho} \right\} +\text{tr}\left\{ \tilde{H} \frac{d\tilde{\rho}}{dt} \right\}.
\end{equation}
In the Schr\"{o}dinger picture this becomes  $\frac{d}{dt} \langle H \rangle = P + \dot{Q}$, where $\dot{Q}$ is the heat current and $P$ is the power which are respectively given by \cite{Alicki_1979}
\begin{equation} \label{eq:Alicki}
	\dot{Q} = \text{tr}\left\{ H \frac{\partial \rho}{\partial t} \right\} \quad \text{and} \quad P = \text{tr} \left\{ \frac{\partial H}{\partial t} \rho \right\}.
\end{equation}
Following the above definitions, if we define heat currents and power in an arbitrary frame as 
\begin{equation} 
	\dot{\tilde{Q}} = \text{tr} \left\{ \tilde{H} \frac{d\tilde{\rho}}{dt} \right\} \quad \text{and} \quad \tilde{P} = \text{tr} \left\{ \frac{d \tilde{H}}{dt} \tilde{\rho} \right\},
\end{equation}
then we have
\begin{subequations}
	\label{eq:rotated_all}
	\begin{eqnarray} 
		\dot{\tilde{Q}} &=& \text{tr}\left\{ \tilde{H} \frac{d\tilde{\rho}}{dt} \right\} = i\; \text{tr} \left\{ [H, x ] \rho \right\} + \text{tr}\left\{\frac{\partial \rho}{\partial t} H\right\} \label{eq:rotated_heat}, \\
		\tilde{P} &=& \text{tr}\left\{ \frac{d \tilde{H}}{dt} \tilde{\rho} \right\} = -i\; \text{tr}\left\{ [H, x ] \rho \right\} + \text{tr}\left\{\frac{\partial H}{\partial t} \rho\right\} \label{eq:rotated_power}.
	\end{eqnarray}
\end{subequations}
From Eq.~(\ref{eq:Alicki}) and Eq.~(\ref{eq:rotated_all}), we can see that the heat current and power are frame dependent.
Authors of reference \cite{boukobza2007three} pointed out a further complication and showed that at certain values of detuning Eq.(\ref{eq:Alicki}) violates the second law of thermodynamics.
To circumvent these problems, Boukobza and Tannor considered the system energy alone and redefined the heat currents and power in terms of the energy currents of the bare Hamiltonian~\cite{boukobza2006thermodynamics}.
The energy flux for the system alone is given by
\begin{eqnarray}
	\frac{d}{dt} \text{tr}\left\{\rho H_0 \right\} & =& \text{tr} \left\{ \frac{d\rho}{dt}H_0 \right\}\nonumber\\
	&=& -i \; \text{tr}\left\{ [H, \rho] H_0 \right\}+ \text{tr}\left\{ \mathcal{L}[\rho] H_0\right\},
\end{eqnarray}
where the first term is due to the Hamiltonian part of the evolution which is defined as the power and the second term is due to the dissipation part of the evolution which is defined as the heat current. 


\emph{Steady States, Heat and Work for a 3-level Quantum Engines.---}	
The Hamiltonian describing our system in the Schr\"{o}dinger picture can be written as a sum of the bare Hamiltonian and a time-dependent drive,
\begin{equation}
	H = H_0 + V(t).
\end{equation}
The drive $V(t)$ has the following form, 
\begin{equation}
	V(t) = \varepsilon \left(\sigma_{23}e^{i\omega_d t} + \sigma_{32} e^{-i\omega_d t}\right),
\end{equation}
where $\sigma_{ij} \equiv |i\rangle\langle j|$.
The Hamiltonian may be written in time-independent form by transforming into a rotating frame via unitary,
\begin{equation}
	U(t) = e^{-i\omega_1t}\sigma_{11} + e^{-i\omega_2t}\sigma_{22}+ e^{-i(\omega_2+\omega_d)t}\sigma_{33}.
\end{equation}
In the rotating frame the Lindblad master equation becomes
\begin{equation}
	\label{eq:master_equation}
	\dot{\tilde{\rho}} = -i [\tilde{H}, \tilde{\rho}] + \mathcal{L}_h[\tilde{\rho}] + \mathcal{L}_c[\tilde{\rho}],
\end{equation}
with
\begin{subequations}
	\begin{eqnarray}
		\tilde{H} &=& \Delta\sigma_{33} + \varepsilon \left( \sigma_{23} + \sigma_{32} \right), \\
		\mathcal{L}_h[\tilde{\rho}] &=& \gamma_h(\bar{n}_h + 1) \mathcal{D} [\sigma_{13}] \tilde{\rho} + \gamma_h \bar{n}_h \mathcal{D} [\sigma_{31}] \tilde{\rho}, \\
		\mathcal{L}_c[\tilde{\rho}] &=& \gamma_c(\bar{n}_c + 1) \mathcal{D} [\sigma_{12}] \tilde{\rho} + \gamma_c \bar{n}_c \mathcal{D} [\sigma_{21}] \tilde{\rho},
	\end{eqnarray}
\end{subequations}
where $\Delta\equiv\omega_{32} - \omega_d$.
The equations of motion for the density matrix elements are
\begin{subequations}
	\begin{eqnarray}
		\dot{\tilde{\rho}}_{11} &=& \gamma_h ( \bar{n}_h + 1) \tilde{\rho}_{33} + \gamma_c ( \bar{n}_c + 1 ) \tilde{\rho}_{22} - (\gamma_h \bar{n}_h + \gamma_c \bar{n}_c) \tilde{\rho}_{11}, \\
		\dot{\tilde{\rho}}_{22} &=& i\varepsilon ( \tilde{\rho}_{23} - \tilde{\rho}_{32} ) - \gamma_c ( \bar{n}_c + 1 ) \tilde{\rho}_{22} + \gamma_c\bar{n}_c \tilde{\rho}_{11}, \\
		\dot{\tilde{\rho}}_{33} &=& -i\varepsilon ( \tilde{\rho}_{23} - \tilde{\rho}_{32} ) - \gamma_h ( \bar{n}_h + 1 ) \tilde{\rho}_{33} + \gamma_h\bar{n}_h \tilde{\rho}_{11}, \\
		\dot{\tilde{\rho}}_{12} &=& -\Gamma_{12} \tilde{\rho}_{12} + i\varepsilon \tilde{\rho}_{13}, \\
		\dot{\tilde{\rho}}_{13} &=& -\Gamma_{13} \tilde{\rho}_{13} + i\varepsilon \tilde{\rho}_{12}, \\
		\dot{\tilde{\rho}}_{23} &=& -\Gamma_{23} \tilde{\rho}_{23} - i\varepsilon \left( \tilde{\rho}_{33} - \tilde{\rho}_{22} \right),
	\end{eqnarray}
\end{subequations}
where the dissipation rates are defined by
\begin{subequations}
	\begin{eqnarray}
		\Gamma_{12} &\equiv& \frac{1}{2} \left[ \gamma_h\bar{n}_h + \gamma_c (2\bar{n}_c + 1) \right], \\
		\Gamma_{13} &\equiv& \frac{1}{2} \left[ \gamma_h (2\bar{n}_h  + 1) + \gamma_c\bar{n}_c \right] - i\Delta, \\
		\Gamma_{23} &\equiv& \frac{1}{2} \left[ \gamma_h(\bar{n}_h + 1) + \gamma_c(\bar{n}_c + 1) \right] - i\Delta.
	\end{eqnarray}
\end{subequations}
Solving for the steady state by setting $\dot{\tilde{\rho}}_{ij} =0$, we get
\begin{equation}
	\tilde{\rho}^{ss}_{11} = \frac{\alpha_1}{\beta}, \qquad \tilde{\rho}^{ss}_{22} = \frac{\alpha_2}{\beta}, \qquad \tilde{\rho}^{ss}_{33} = \frac{\alpha_3}{\beta}, \qquad \tilde{\rho}^{ss}_{23} = \frac{\alpha_4}{\beta},
\end{equation}
with the remaining coherences vanishing, $\tilde{\rho}^{ss}_{12}=\tilde{\rho}^{ss}_{13}=0$, and where
\begin{subequations}
	\label{eq:rhoss_solution}
	\begin{eqnarray}
		\alpha_1 &=& 2\epsilon ^2 \text{Re} [\Gamma_{23}] \left[ \gamma_c (\bar{n}_c + 1) + \gamma_h (\bar{n}_h + 1) \right] + |\Gamma_{23}|^2 \gamma_c \gamma_h (\bar{n}_c+1) (\bar{n}_h+1), \\
		\alpha_2 &=& 2\epsilon ^2 \text{Re} [\Gamma_{23}] (\gamma_c \bar{n}_c+\gamma_h \bar{n}_h)+ |\Gamma_{23}|^2 \gamma_c \gamma_h \bar{n}_c (\bar{n}_h+1), \\
		\alpha_3 &=& 2\epsilon ^2 \text{Re} [\Gamma_{23}] (\gamma_c \bar{n}_c+\gamma_h \bar{n}_h)+ |\Gamma_{23}|^2 \gamma_c \gamma_h (\bar{n}_c+1) \bar{n}_h, \\
		\alpha_4 &=& i  \Gamma_ {23}^* \gamma_c \gamma_h \epsilon  (\bar{n}_c-\bar{n}_h), \\
		\beta &=& 2\epsilon ^2 \text{Re} [\Gamma_{23}] \left[ \gamma_c (3\bar{n}_c + 1) + \gamma_h (3\bar{n}_h + 1) \right]+ |\Gamma_{23}|^2 \gamma_c \gamma_h \left[ \bar{n}_c (3 \bar{n}_h+2)+2 \bar{n}_h+1 \right].
	\end{eqnarray}
\end{subequations}
The power and heat currents in the steady state are
\begin{subequations}
	\begin{eqnarray}
		P^{ss} &\equiv& -i \;\text{tr} \left\{ [\tilde{H}, \tilde{\rho}^{ss}] H_0 \right\} = 2 \varepsilon \omega_{32} \text{Im} \left[ \tilde{\rho}^{ss}_{23} \right], \\
		\dot{Q}^{ss}_h &\equiv& \text{tr} \left\{ \mathcal{L}_h[\tilde{\rho}^{ss}]H_0 \right\} = \gamma_h\bar{n}_h\omega_{31} (\tilde{\rho}^{ss}_{11} - \tilde{\rho}^{ss}_{33}) - \gamma_h\omega_{31}\tilde{\rho}^{ss}_{33}, \\
		\dot{Q}^{ss}_c &\equiv& \text{tr} \left\{ \mathcal{L}_c[\tilde{\rho}^{ss}]H_0 \right\} = \gamma_c\bar{n}_c\omega_{21} (\tilde{\rho}^{ss}_{11} - \tilde{\rho}^{ss}_{22}) - \gamma_c\omega_{21}\tilde{\rho}^{ss}_{22}.
	\end{eqnarray}
\end{subequations}
Using Eq.~(\ref{eq:rhoss_solution}) the heat currents can be rewritten,
\begin{subequations}
	\begin{eqnarray}
		\dot{Q}^{ss}_h &=& \frac{2\gamma_h\gamma_c\varepsilon^2\omega_{31}\text{Re}[\Gamma_{23}] (\bar{n}_h - \bar{n}_c)}{\beta}, \\
		\dot{Q}^{ss}_c &=& -\frac{2\gamma_h\gamma_c\varepsilon^2\omega_{21}\text{Re}[\Gamma_{23}] (\bar{n}_h - \bar{n}_c)}{\beta}.
	\end{eqnarray}
\end{subequations}
We see that for $\bar{n}_h>\bar{n}_c$ we have $\dot{Q}^{ss}_h>0$ and $\dot{Q}^{ss}_c<0$ meaning the energy flows from the hot bath to the cold bath and the system behaves like an engine.
On the other hand when $\bar{n}_h<\bar{n}_c$ the signs of the heat currents and power reverse and the system behaves like a fridge.

\end{document}